\documentclass[conference]{IEEEtran}

\usepackage{listings}
\usepackage{color}
\usepackage{makecell}
\usepackage{float}

\usepackage{graphicx}

\definecolor{red}{rgb}{1, 0, 0}
\definecolor{dkgreen}{rgb}{0,0.6,0}
\definecolor{gray}{rgb}{0.5,0.5,0.5}
\definecolor{mauve}{rgb}{0.58,0,0.82}

\begin{document}

\title{
Link Prediction Using Supervised Machine Learning based on Aggregated and Topological Features
}

\author{\IEEEauthorblockN{Mohammad G. Raeini}\\
\IEEEauthorblockA{
Department of CEECS, Florida Atlantic University
}
}

\maketitle

\begin{abstract}
Link prediction is an important task in social network analysis. There are different characteristics (features) in a social network that can be used for link prediction. In this paper, we evaluate the effectiveness of aggregated features and topological features in link prediction using supervised learning. The aggregated features, in a social network, are some aggregation functions of the attributes of the nodes. Topological features describe the topology or structure of a social network, and its underlying graph.

We evaluated the effectiveness of these features by measuring the performance of different supervised machine learning methods. Specifically, we selected five well-known supervised methods including J48 decision tree, multi-layer perceptron (MLP), support vector machine (SVM), logistic regression and Naive Bayes (NB). We measured the performance of these five methods with different sets of features of the DBLP Dataset. Our results indicate that the combination of aggregated and topological features generates the best performance.
For evaluation purposes, we used accuracy, area under the ROC curve (AUC) and F-Measure.

Our selected features can be used for the analysis of almost any social network. This is because these features provide the important characteristics of the underlying graph of the social networks. The significance of our work is that the selected features can be very effective in the analysis of big social networks. In such networks we usually deal with big data sets, with millions or billions of instances. Using fewer, but more effective, features can help us for the analysis of big social networks.
\end{abstract}

\IEEEpeerreviewmaketitle
\section{Introduction}
Internet is growing at an extremely rapid pace which in effect has greatly improved the number of individuals communicating and collaborating with each other. These interactions have become effortless due to the advancements of popular information exchange platforms, e.g. online social networks. Social networks can be visualized as graphs between multiple nodes. In such graphs the nodes represent entities, e.g. people or organizations, linked together based on different factors, e.g. friendship. Due to the increase of social networks, social network analysis has attracted significant attention in recent years.

Social Network Analysis (SNA) is a broad field of research dealing with techniques and strategies for the study and analysis of social networks \cite{SaPrudencio2011}. SNA allows us to analyze and determine the relations between the nodes in the social network. In particular, link prediction is an important task that is used vastly in social network analysis \cite{SaPrudencio2011}. Link prediction is a method used to pre-determine connection between nodes by using the current patterns observed from the network. These observations can help to better understand the future changes both in the graph and between individual nodes.
Link prediction is used in many areas including information retrieval, bioinformatics, e-commerce, criminal investigations and recommender systems \cite{AhmedElKoranyBahgat2016}.

In social networks, link prediction can be used to predict the topics a user would be interested in. The prediction is based on the user's previous likes and dislikes of the social network content. Another example wherein link prediction has been used extensively is e-commerce and online shopping.  In this case, link prediction is used to offer the user the items they might be interested in. The prediction in this case relies on the user's previous purchases, key terms they searched and other activities. Link prediction is also used to determine the likelihood of association between the authors of co-authorship social networks.

In this paper, we specifically study link prediction in co-authorship social networks.
We use supervised machine learning methods on aggregated features and topological features.
Our goal is to evaluate the effectiveness of these features for link prediction in such social networks. We selected the DBLP dataset of the co-authorship social network. 
This dataset contains detailed information of the scientific papers published in computer science since 1993. We extracted the above-mentioned features for the instances of this dataset. We then used five different supervised machine learning techniques to evaluate the effectiveness of the selected features for the link prediction problem. The selected machine learning techniques include J48 decision tree, support vector machines (SVM), Multi-Layer Perceptron (MLP), Logistic regression and Naive Bayes (NB).

The rest of this project is organized as follows. In section \ref{related-works}, we review the previous works related to link prediction in social networks. In section \ref{main-body}, we discuss the supervised machine learning methods and the dataset that we use for the link prediction. We also explain the performance measures that we will use for measuring the performance of the selected methods.  In section \ref{experimental-results}, the experimental study as well as the analysis of the results will be provided. In section \ref{conclusion}, we conclude the paper and summarize our results.

\section{Related Works}
\label{related-works}
Machine learning algorithms are appropriate tools for analyzing and predicting the behavior of entities in a social network. So far, there have been different solutions for the link prediction problem in social networks. The previous works \cite{BenchettaraKanawatiRouveirol2010, DavisLichtenwalterChawla2012, BenchettaraKanawatiRouveirol2010InternationalConference, BackstromLeskovec2011, MurataMoriyasu2008} that we review in this section mostly used supervised machine learning methods for link prediction in different ways and circumstances.

In \cite{BenchettaraKanawatiRouveirol2010}, predicting future possible co-authorship was studied as a link prediction problem. The researchers \cite{BenchettaraKanawatiRouveirol2010} suggested that co-authorship network is actually a projected graph: a projection of a bipartite graph linking authors to publications \cite{BenchettaraKanawatiRouveirol2010}. Knowing this, the prediction problem was converted into linking or non-linking class discrimination problem. The class discrimination problem, in turn, was further explained in \cite{HasanChaojiSalemZaki2006, BenchettaraKanawatiRouveirol2010, MurataMoriyasu2008}. This approach permits for the problem to be separated into two class discrimination problems. Two class discrimination problems can be solved using classical machine learning techniques. This approach showed it works well on the computer science bibliography dataset.

The Dyadic method is another well-known approach for solving the link prediction problem. This method works by assigning a value (score) to each small group of unlinked vertices.  Many of the link predication methods apply a dyadic approach. Systems use different ranking functions such as topological score or node features \cite{BenchettaraKanawatiRouveirol2010, LibenNowell2005}. Structural approaches, however, search the whole network for mining rules of evolution of subgraph; which can in return predict the appearance or disappearance of multiple links at once.

Node-based and topology-based methods are some other common approaches for link prediction. To find future links between nodes, their similarity can be used as key factor. This is because the more similar two nodes are, the more likely it is for them to form a link \cite{WangXuWuZhou2014, YangLichtenwalterChawla2014}. For Nodes in the academic social networks, this factor is calculated using the publication records inside the network. Topology, however, can be more successful due to it being application dependent. In the Topological method, certain features such as Common Neighbors, Jaccard’s Coefficient, Sum of Neighbors, etc. are extracted and used to find the similarity score between the nodes in the network.

Supervised random walks is another approach that can be used for both link recommendation and link prediction. It is done by combining the structure of the network with the nodes' features and attributes to create a unified link prediction algorithm. The algorithm can then be evaluated by preforming supervised PageRank like bias walk on the network so that it can determine whether a node is positive (new edges will be created) or negative (all other nodes).

\section{Supervised Link Prediction using Aggregated and Topological Features}
\label{main-body}
In this paper, we evaluate the effectiveness of aggregated and topological features for link prediction in social networks. We use five well-known supervised machine learning methods and measure their performance on the DBLP co-authorship social network. These methods include J48 decision tree, support vector machines (SVM), Multi-Layer Perceptron (MLP), Logistic regression and Naive Bayes (NB).

For the DBLP dataset, the aggregated features that we consider are \textit{sum of papers} and \textit{sum of neighbors}. The considered topological feature is the \textit{shortest distance}, between pairs of the nodes of the social network's graph. Note that aggregated and topological features are intrinsic characteristics of the underlying graph of any social network. Our goal is to see how effective these features are in terms of predicting the future links in a social network.

We extract the aforementioned features from the DBLP dataset and construct the training and test data sets. We then train and evaluate the five selected machine learning methods on the constructed data sets. For performance measures, we use accuracy, area under the ROC curve (AUC), and F-measure.

\subsection{Supervised Machine Learning Methods}
\label{supervised-learning-methods}
In this section, we briefly review the supervised machine learning techniques that we use in our experiments.

\textit{J48 Decision Tree (J48):}
J48 decision tree is one of the most effective machine learning algorithms for classification \cite{PatilSherekar2013}. This algorithm in its essence is a simple C4.5 decision tree \cite{PatilSherekar2013}. It works by creating a binary tree and applying the classification rules to each instance of the training data set and returning the results as a decision tree. J48 is efficient due to the fact that it skips the tuples which are missing values and attempts to predict the value based on the information it gathers from others that share the same range of attributes.

\textit{Support Vector Machines (SVM):}
SVM or Support Vector Machine is one of the most popular and widely used algorithms in machine learning and data mining. It operates by iterating through the training data set. This algorithm creates an ideal hyperplane, which is a hypothetical line that divides the input variable space. The hyperplane helps in classifying the instances of the data set accurately.

\textit{Multi-Layer Perceptron (MLP):}
Multi-Layer Perceptron, or MLP for short, is an artificial neural network consisted of some layers of neurons. MLP is considered as a supervised machine learning algorithm. This neural network is trained using a technique known as backpropagation, which uses the training data.
After training, the network can predict the class of new instances. In order to do that, the new instance(s) are fed into the network and the network classifies the instance.

\textit{Logistic Regression (Log-Reg):}
Logistic Regression is a well-known statistical analysis method, which is also used for classification problems \cite{PopesculUngar2003}. It is a good model to describe the data and explain the relationship between a dependent variable and some independent variables. Logistic regression is an appropriate method for data analysis when the dependent variable is a binary variable. In terms of classification problems, logistic regression models the conditional class probabilities with no regards to modeling the marginal distribution of features \cite{PopesculUngar2003}.

\textit{Naive Bayes (NB):}
Naive Bayes is another well-known classification method. It is based on the Bayes theorem and assumes all the attributes are independent, given that the class variable is specified \cite{PatilSherekar2013}. Naive Bayes counts the frequency and combinations of the values in a given data set and then calculates a set of probabilities. Using the calculated probabilities, it finds the class of new instances. The conditional independence of attributes is unrealistic and rarely holds \cite{ZafaraniAbbasiLiu2014}. However, this condition simplifies the calculations significantly. It also lets Naive Bayes to perform well.

\subsection{The Dataset}
\label{the-dataset}

\begin{figure}[H]
	\begin{center}
		\includegraphics[scale=0.5]{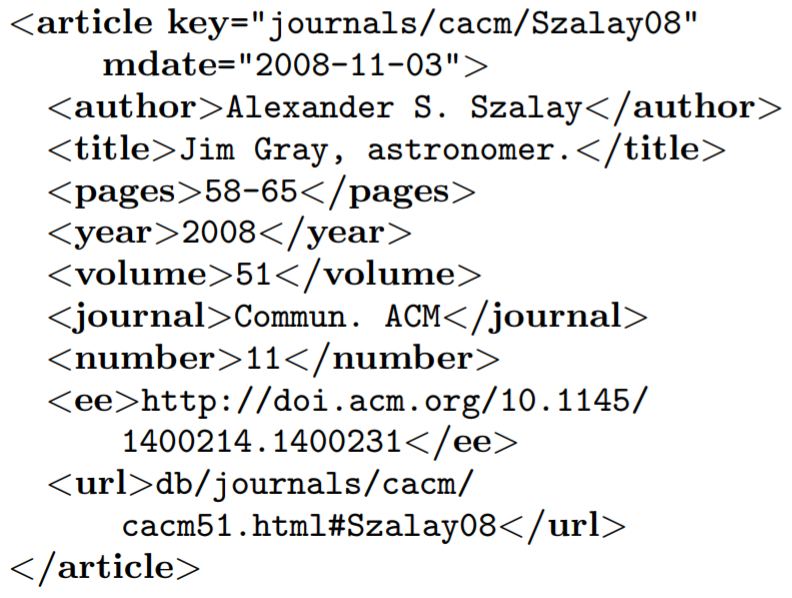}
	\end{center}
	\caption{A sample record of the DBLP dataset \cite{dblp2009}}
	\label{fig:sample-record-of-dblp-dataset}
\end{figure}

We used the DBLP dataset in our experiments to evaluate the performance of our models.
DBLP is a dataset of the bibliography of scientific publications in the field of computer science since 1993 \cite{dblp2009}. This data set contains the details of the publications including the title, the authors, the journal and other details of publications. Originally at the University of Trier in Germany, it was meant to index scientific works in the areas of database and logic programming, which suggested its name as DBLP. However, now it includes all publications of the computer science field. This dataset is a huge XML file (about 2 Giga bytes), containing more than 3.5 million publications. It is available for download at \cite{dblpDownloadLink}. According to \cite{dblp2009}, this data set includes different attributes such as the authors, the title, the pages, year, Crossref, URL and EE, for each record, i.e. publication. However, the data set does not have keywords related to any record. Figure \ref{fig:sample-record-of-dblp-dataset} shows a sample record (instance) of this data set.

\subsection{Feature Selection}
\label{subsec:feature-selection}
We defined two types of features, aggregated and topological features, to be used in the machine learning methods. From the aggregated type of features, we selected two features, i.e. the \textit{sum of neighbors} and the \textit{sum of papers}. From the topological type of features, we selected only one feature, i.e. the \textit{shortest path distance} feature. In what follows, we briefly explain these features.

\begin{itemize}
  \item \textit{Shortest path distance:} the shortest path distance is one of the most important features in link prediction \cite{HasanChaojiSalemZaki2006}.
Our experimental results also confirms this fact. This is because, according to \cite{HasanChaojiSalemZaki2006}, in social networks the distance between most of the nodes is very short. So, the shortest distance is a good feature for the link prediction problem. In fact, it is obvious that the closer the nodes are, the more likely they will form a new connection (link). Recall that this feature falls in the topological features category.
  \item \textit{Sum of neighbors:} the sum of neighbors indicates the total number of neighbors each pair of authors have. For example, if two authors have 3 and 4 neighbors then sum of neighbors would be 7. For this feature if two authors have some common neighbors, we count them once. The reason that this feature is considered is that if a node is connected to many other nodes, then it is more likely this node build new connection (link). We consider this feature as an aggregated feature, although it can be considered as a topological feature \cite{HasanChaojiSalemZaki2006}.
  \item \textit{Sum of Papers: } for calculating the value of this feature we simply add the number of papers that the authors have published. This feature is also considered as an aggregated feature. The reason for selecting this feature is that authors who have many publication are more likely to create new links.
\end{itemize}

Besides these three features, each instance (record) of our dataset has a class attribute which indicates the class of that instance. The class label (value) can be 1 or 0, which shows the two authors are connected or not respectively.

After preparing the datasets, we used 5 supervised learning models (implemented in Weka software) to train them and then test them on the test dataset. All the experimental results have been included in the experimental results section (section \ref{experimental-results}).

\subsection{Evaluation Measures}
\label{evaluation-measures}
The performance measures that we use for the evaluation of our models are as follows.

\textit{Accuracy:}
Accuracy is a common measurement for evaluating the performance of machine learning methods. It provides us with a measure to compare the performance of different methods.
In other words, it allows us to see which method has a better yield, or provides better results. Accuracy is defined as the ratio of correctly classified instances out of the total number of instances in a given data set. Mathematically speaking, accuracy is defined as in equation \ref{eq:accuracy}.

\begin{equation}
\label{eq:accuracy}
Accuracy = \frac{TP + TN}{TP + TN + FP + FN}
\end{equation}

where TP (True Positive) is the number of instances correctly classified as positive, TN (True Negative) is the number of instances correctly classified as negative. FP (False Positive) indicates the number of instances that have been incorrectly classified as positive, and FN (False Negative) shows the number of instances that have been incorrectly classified as negative.

\textit{Area Under ROC Curve (AUC):}
The next performance measurement that we use in this paper is Area Under ROC Curve (AUC). Receiver operating characteristic curve (ROC curve) displays the performance of a certain classification model at all different possible thresholds using TPR (True Positive Rate) and FPR (False Positive Rate). AUC, however, is a more efficient criteria for calculating the full two-dimensional area beneath the ROC curve \cite{Powers2011}. AUC can provide the overall performance of a model at all different possible classification thresholds, rather than at specific thresholds. The area under the ROC curve (AUC) varies between zero and one.
A model performs better than chance if its AUC is more than 0.5. In general, the higher the AUC, the better the model performs.

\textit{F-Measure:}
F-Measure is another measure for evaluating the performance of a model. It is considered to be the harmonic mean of the precision and recall. Precision is the proportion of truly classified instances as positive (TP), amongst all the positive instances.

\begin{equation}
Precision = \frac{TP}{TP + FP}
\end{equation}

Recall, however, tries to uncover the proportion of True Positives that were correctly identified.

\begin{equation}
Recall = \frac{TP}{TP + FN}
\end{equation}

F-Measure computes the accuracy of a model by taking a harmonic mean of the \textit{precission} and \textit{recall} measures as follows.

\begin{equation}
F1 = \frac{2}{\frac{1}{recall} + \frac{1}{precision}}
\end{equation}

\section{Data Preparation and Experimental Results}
\label{experimental-results}
In this section, we explain how we prepared our data and extracted the desired features. We also detail our experiments and analyze our results. The supervised machine learning methods, the DBLP dataset, the features and the performance measures that we use in this paper were explained in sections \ref{supervised-learning-methods}, \ref{the-dataset}, \ref{subsec:feature-selection} and \ref{evaluation-measures}, respectively.

\subsection{Data preparation}
In our study, we treat the social network as a graph $G = \langle V, E \rangle$ where $V$ is the set of nodes and $E$ is the set of edges. The data set that we use in our experiments is the DBLP data set. In this dataset, the set of nodes $G$ contains the authors of the publications. If two authors have co-authored a publication, then there is an edge in the set of edges $E$. In our implementation we represented the social network as a graph, using the graph's adjacency matrix. We assigned each author a unique ID, starting from 1, 2 and so forth. Therefore, we have a graph $G = \langle V, E \rangle$ where $G = \lbrace 1, 2, ... \rbrace$. If authors $i$ and $j$ have published a paper, then there is an edge $(i, j)$ connecting them. The rows and columns of the matrix are the IDs of the authors. If authors $i$ and $j$ have co-authored a paper, then the corresponding element of the matrix is 1, otherwise 0. That is, if authors $i$ and $j$ are co-author, then the element of the adjacency matrix at $(i, j)$ is 1.

The DBLP data set is very big; it indexed more than 3.5 million publications and its file is about 2 Gigabytes in size. Since it was not possible to process it as a single file, due to the limitations of Java JDK, we only selected small portion of this data set. To be more specific, we used the first 50000 lines of this data set. We then selected the papers published between years 2012 and 2018. Among these papers, we chose those paper published in years between 2012 and 2016 ($2012 \leq year \leq 2016$) to be included in our training data set. We also constructed a test data set of those papers published in 2017 and 2018 ($2017 \leq year \leq 2018$). Note that the experiment was done in 2018. So, we considered the papers of the early months of 2018.

Table \ref{table:detaset-for-experiment} shows the details of the data set we used for our experiments. In total our training data set contains 1101 paper and 3376 authors. Our test data set includes 1107 authors and 322 papers.

\begin{table}[H]
\begin{center}
\begin{tabular}{ |c||c|c|  }
 \hline
 \multicolumn{3}{|c|}{Our data sets extracted from the DBLP data set} \\
 \hline
 Dataset & Number of Papers & Number of Authors\\
 \hline
 Training   & 1101    & 3376\\
 \hline
 Test   & 322    & 1107\\
 \hline
\end{tabular}
\caption{The data sets used for the experiment
\label{table:detaset-for-experiment}}
\end{center}
\end{table}

For any connected pair of nodes (authors) in the network, we defined two aggregated features, i.e. the \textit{sum of neighbors} and the \textit{sum of papers}. For the \textit{sum of neighbors}, we need to count all the neighbors of any pair of authors who are co-author.
For the \textit{sum of papers}, we need to add up the number of the papers that have been published by the two authors. We also defined a topological feature, i.e. \textit{the shortest path distance}, between any pair of authors. For finding the shortest path distance, we used the Dijkstra's algorithm, implemented in Java.

\begin{figure}[H]
	\begin{center}
		\includegraphics[scale=0.7]{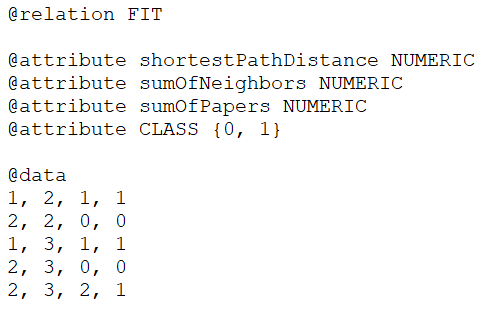}
	\end{center}
	\caption{Sample instances of our constructed dataset}
	\label{fig:sample-arff-file}
\end{figure}

In the next phase, we created our classification data sets (ARFF files) to be used in the Weka software. To do so, we extracted two types of features: aggregated features (sum of neighbors and sum of papers) and topological features (shortest path). Moreover, we assigned a class to each instance in such a way that if the two nodes are connected (are co-author ) we assigned the class 1 and if they are not connected the class is 0. Therefore, our problem is a binary classification. Figure \ref{fig:sample-arff-file} shows some instances of our constructed dataset.

\subsection{Experimental Results}

In this section, we explain how we performed our experiments. Our goal is to evaluate the effectiveness of the aggregated and topological features for link prediction in social networks. In other words, we want to see how well a machine learning model can perform, where the model is trained based on aggregated and topological features.

After preparing our data and extracting our favorite features, we used five different well-known supervised machine learning techniques for the link prediction problem. The techniques we selected include J48 decision tree, support vector machines (SVM), Multi-Layer Perceptron (MLP), logistic regression (in the tables we called it Log-Reg, due to space constraints) and Naive Bayes. For stability of the results we used 10-fold cross validation, which is the default in the Weka software. For evaluation purposes, we used accuracy, area under ROC curve (AUC) and F-Measure.

As mentioned earlier we used three features in our experiment. These features include two aggregated features, i.e the \textit{sum of neighbors} and the  \textit{sum of papers} and one topological feature, i.e. the \textit{the shortest distance} between pairs of nodes (authors). Then we used five different supervised machine learning methods on different combination of features. We measured the performance of the models for each combination.
Table \ref{table:performance-of-different-supervised-learning-techniques} shows the performance of different techniques using only topological feature, i.e. \textit{the shortest distance}.

\begin{table}[H]
\begin{center}
\begin{tabular}{ |c||c|c|c|  }
 \hline
 \multicolumn{4}{|c|}{Performance of different models on the training data set} \\
 \hline
 Model & Accuracy & AUC & F-Measure\\
 \hline
 J48   & 97.65  &  0.976 & 0.976\\
 \hline
 SVM   & 97.65    & 0.978 & 0.976\\
 \hline
  MLP   & 97.65    & 0.989 & 0.976\\
 \hline
  Log-Reg   & 90.13    & 0.847 & 0.900\\
 \hline
  Naive Bayes & 97.65 & 0.988 & 0.976\\
 \hline
\end{tabular}
\caption{Performance of the models on the training data set using \textit{the shortest distance} feature
\label{table:performance-of-different-supervised-learning-techniques}}
\end{center}
\end{table}

The performances of different models on our test data set are summarized in the table \ref{table:performance-of-different-supervised-learning-techniques-test-dataset}.

\begin{table}[H]
\begin{center}
\begin{tabular}{ |c||c|c|c|  }
 \hline
 \multicolumn{4}{|c|}{Performance of different models on the test data set} \\
 \hline
 Model & Accuracy & AUC & F-Measure\\
 \hline
 J48   & 97.99  &  0.989 & 0.918\\
 \hline
 SVM   & 97.99    & 0.989 & 0.918\\
 \hline
  MLP   & 97.99    & 0.990 & 0.918\\
 \hline
  Log-Reg   & 97.99    & 0.990 & 0.918\\
 \hline
  Naive Bayes & 97.99 & 0.990 & 0.918\\
 \hline
\end{tabular}
\caption{Performance of the models on test data set using \textit{the shortest distance} feature
\label{table:performance-of-different-supervised-learning-techniques-test-dataset}}
\end{center}
\end{table}

Next, we investigated the performances of the models using one of the aggregated features, i.e. the \textit{sum of neighbors}. Tables \ref{table:performance-of-different-supervised-learning-techniques-sumOfNeighbors} and \ref{table:performance-of-different-supervised-learning-techniques-test-dataset-sumOfNeighbors} show the performances of different models on training and test data sets respectively.

\begin{table}[H]
\begin{center}
\begin{tabular}{ |c||c|c|c|  }
 \hline
 \multicolumn{4}{|c|}{Performance of different models on the training data set} \\
 \hline
 Model & Accuracy & AUC & F-Measure\\
 \hline
 J48   & 77.97  &  0.786 & 0.779\\
 \hline
 SVM   & 78.06    & 0.779 & 0.764\\
 \hline
  MLP   & 78.10    & 0.838 & 0.783\\
 \hline
  Log-Reg   & 78.06    & 0.843 & 0.764\\
 \hline
 Naive Bayes & 74.25 & 0.844 & 0.701\\
 \hline
\end{tabular}
\caption{Performance of the models on the training data set using the \textit{sum of neighbors} feature
\label{table:performance-of-different-supervised-learning-techniques-sumOfNeighbors}}
\end{center}
\end{table}

\begin{table}[H]
\begin{center}
\begin{tabular}{ |c||c|c|c|  }
 \hline
 \multicolumn{4}{|c|}{Performance of different models on the test data set} \\
 \hline
 Model & Accuracy & AUC & F-Measure\\
 \hline
 J48   & 76.31  &  0.664 & 0.337\\
 \hline
 SVM   & 85.45    & 0.663 & 0.391\\
 \hline
  MLP   & 76.31    & 0.774 & 0.337\\
 \hline
  Log-Reg   & 85.45    & 0.774 & 0.391\\
 \hline
 Naive Bayes & 86.34 & 0.774 & 0.280\\
 \hline
\end{tabular}
\caption{Performance of the models on the test data set using the \textit{sum of neighbors} feature
\label{table:performance-of-different-supervised-learning-techniques-test-dataset-sumOfNeighbors}}
\end{center}
\end{table}

\begin{figure}[H]
	\begin{center}
		\includegraphics[scale=0.7]{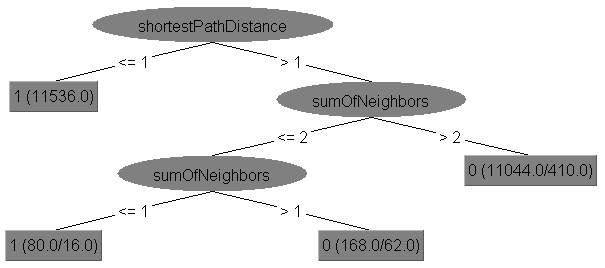}
	\end{center}
	\caption{Constructed decision tree using both features on the training dataset}
	\label{fig:j48-tree}
\end{figure}

Finally, we use both type of the features, i.e. the aggregated and the topological features, to see the performances of the models. The results are shown in figure 
\ref{table:performance-of-different-supervised-learning-techniques-both-features} for the training data set and figure 
\ref{table:performance-of-different-supervised-learning-techniques-test-dataset-both-features} for the test data set. Figure \ref{fig:j48-tree}, for instance, shows the constructed decision tree using these features on the training data set.

\begin{table}[h!]
\begin{center}
\begin{tabular}{ |c||c|c|c|  }
 \hline
 \multicolumn{4}{|c|}{Performance of different models on the training data set} \\
 \hline
 Model & Accuracy & AUC & F-Measure\\
 \hline
 J48   & 98.86  &  0.989 & 0.978\\
 \hline
 SVM   & 97.65  &  0.978 & 0.976\\
 \hline
  MLP   & 97.81  &  0.993 & 0.977\\
 \hline
  Log-Reg  & 97.65  &  0.993 & 0.976\\
 \hline
  Naive Bayes & 96.53 & 0.991 & 0.964\\
 \hline
\end{tabular}
\caption{Performance of the models on the training data set using the \textit{sum of neighbors} and the \textit{shortest distance} features
\label{table:performance-of-different-supervised-learning-techniques-both-features}}
\end{center}
\end{table}

\begin{table}[h!]
\begin{center}
\begin{tabular}{ |c||c|c|c|  }
 \hline
 \multicolumn{4}{|c|}{Performance of different models on the test data set} \\
 \hline
 Model & Accuracy & AUC & F-Measure\\
 \hline
 J48   & 97.85  &  0.989 & 0.912\\
 \hline
 SVM   & 97.99  &  0.989 & 0.918\\
 \hline
  MLP   & 97.80  &  0.993 & 0.909\\
 \hline
  Log-Reg   & 97.99  &  0.993 & 0.918\\
 \hline
 Naive Bayes & 97.99 & 0.993 & 0.918\\
 \hline
\end{tabular}
\caption{Performance of the models on the test data set using the \textit{sum of neighbors} and the \textit{shortest distance} features
\label{table:performance-of-different-supervised-learning-techniques-test-dataset-both-features}}
\end{center}
\end{table}

\subsection{Analysis of the results}
We used different machine learning methods on different combinations of the selected features and measured the performances of the methods. Next, we discuss and analyze the results of our experiments.

Table \ref{table:performance-of-different-supervised-learning-techniques} shows the performance of different supervised machine learning models on the training data set using the only topological feature, i.e. the \textit{shortest distance}. As we see, this feature is a good one for predicting the future co-authorship of the nodes (authors) in the social network. All the models, except the logistic regression model, have a performance of 97.65\%, which is significantly larger than that of the logistic regression model. The logistic regression model's performance is 90.13\%. However, the models have different AUC values. The MLP model has the highest AUC value of 0.989 and the logistic regression model has the lowest AUC value of 0.847.

The similarity of the models' performances is well explained by their equal F-Measure values. According to \cite{HasanChaojiSalemZaki2006}, F-measure is a balanced mean of precision and recall. Sometimes, it is preferred to accuracy as a performance measure, specifically when the class distribution is biased in our data sets. This observation is in accordance with the findings in \cite{HasanChaojiSalemZaki2006}. The main difference between their experiments and our first experiment is that they selected different features, but we selected only one feature, i.e. the \textit{shortest distance} feature.

On the test data set, the results are summarized in table \ref{table:performance-of-different-supervised-learning-techniques-test-dataset}. As we see the models have similar accuracy and different AUC values, agian similar F-measure values.

In our second experiment, we investigated how much the aggregated features, i.e. the \textit{sum of neighbors} and the \textit{sum of papers}, can help us in link prediction. First of all, including the \textit{sum of papers} feature did not result in good models. In fact, when we use the \textit{sum of papers} feature in the set of the features, the models use this feature and exclude other features. The reason is that this feature solely determines the class of each instance. That is, when the sum of the papers, for two authors, is not zero obviously those authors are co-author and have a link. On the other hand, the other aggregated feature, i.e. \textit{sum of neighbors}, resulted in better models. Table \ref{table:performance-of-different-supervised-learning-techniques-sumOfNeighbors} and table \ref{table:performance-of-different-supervised-learning-techniques-test-dataset-sumOfNeighbors} show the performances of the different models on training and test data sets respectively.

The tables also show that the performances of the models built on the \textit{sum of neighbors} feature are not as good as the performances of the models built on the \textit{shortest distance} feature. The accuracy of the models built on the \textit{sum of neighbors} feature ranges between 74\% and 78\%. The MLP model has the highest value of 78.10\% and Naive Bayes has the lowest value of 74.25\%. Whereas, the accuracy of the models built on the \textit{shortest distance} feature ranges between 90.13\% and 97.65\%.

Finally, we measured the performance of the models using the combination of all the features, i.e. the \textit{shortest distance} and the \textit{sum of neighbors}. Tables \ref{table:performance-of-different-supervised-learning-techniques-both-features} and \ref{table:performance-of-different-supervised-learning-techniques-test-dataset-both-features} show the performances of the models based on these two features on training and test data sets respectively. In this case, the accuracy of the models are very close, and better than using the features separately. On the training data set, J48 decision tree has the highest accuracy, MLP is in the second rank and Naive Bayes has the lowest accuracy. However, MLP and logistic regression have the highest AUC values. On the test data set, the same models have almost the same accuracy and AUC values. Logistic regression and Naive Bayes have the highest accuracy, AUC and F-measure values. With very negligible difference, J48 decision tree has the lowest accuracy. MLP is amongst the models with a high AUC value, but it has the lowest F-measure of 0.909.

\section{Conclusion}
\label{conclusion}
Link prediction is an important task in social network analysis (SNA). It has applications in other fields such as e-commerce, information retrieval, bioinformatics, criminal investigations and recommendation systems. There are different attributes (features) that can be used for the link prediction. In this paper we evaluated the effectiveness of the aggregated and topological features for supervised link prediction.

We evaluated the performance of five different supervised machine learning methods for link prediction in the DBLP co-authorship data set. The machine learning methods we used include J48 decision tree, SVM, MLP, Logistic Regression and Naive Bayes. We constructed a training data set consisting of the papers published in years between 2012 and 2016. We also used the papers published in 2017 and 2018 years as the test data set. We extracted two aggregated features, i.e. the \textit{sum of neighbors} and the \textit{sum of papers}, and topological features, i.e. the \textit{shortest distance}, for any pair of the authors.

Based on our experiments we observed that the \textit{shortest distance} feature is a very effective feature for link prediction. That is, machine learning models built using this feature can predict the future links with high accuracy. This has been verified in the previous works as well \cite{HasanChaojiSalemZaki2006}. Another effective feature for link prediction is the \textit{sum of neighbors} feature. Using the other aggregated feature, i.e. the \textit{sum of papers}, disaffects the importance of other features. That is, for the DBLP data set, models built using the \textit{sum of papers} feature, use only this feature for link prediction. Such models are not good models. This means that using only the aggregated and topological features allows us to build models that can do link prediction with a high accuracy.

Our models have an accuracy of around 97\%. In general, the MLP model performed very well, although other models had also good accuracies. For the purpose of performance evaluation, we used accuracy, AUC and F-measure. The significance of our work is that we determined the most effective features for link prediction in social networks. This can be very helpful for link prediction in big social network, in which we face big data sets.

\section*{}

\end{document}